\documentclass[leqno,orivec]{llncs}
\usepackage{hyperref}
\newcommand{\mykeywords}{%
  Gr\"{o}bner basis; signature-based algorithms; computational algebra; %
  functional programming; Haskell; type system; formal methods; %
  property-based testing; implementation report%
 }

\hypersetup{pdfproducer={LuaLaTeX},pdfkeywords=\mykeywords,%
psdextra,pdfusetitle,pdfencoding=auto}
\usepackage{color,xcolor,booktabs,array,tabularx}
\usepackage{amssymb}	% required for `\mathbb' (yatex added)
\usepackage{multicol}	% required for `\multicols' (yatex added)
\usepackage{mymacros}
\usepackage{dsfont}
\DeclareMathAlphabet{\mathrsfs}{U}{rsfso}{m}{n}
\renewcommand{\mathscr}[1]{\mathup{\mathrsfs{#1}}}
\usepackage{fixme}
\usepackage[super]{nth}
\usepackage{bm}
\usepackage[backend=biber, style=lncs, doi=false]{biblatex}
\usepackage{fixme}
\usepackage{siunitx}
\usepackage{float}
\usepackage{fancyvrb}
\usepackage[final]{listings}
\usepackage{cleveref}
\usepackage{soul}

\floatstyle{ruled}
\newfloat{flcode}{htbp}{lop}
\floatname{flcode}{Code}
\Crefname{flcode}{Code}{Codes} 
\crefname{flcode}{code}{codes} 
\newcommand{\ovalcode}[3][2pt]{%
  \tikz[baseline=(T.base)]{\node[rounded corners,  outer sep=0pt, inner sep=#1,font={\ttfamily\small}, #2] (T) {#3}}%
}
\def\uley#1{{\underline{\smash{#1}}}}
\definecolor{ltblue}{rgb}{0,0.4,0.4}
\definecolor{dkblue}{rgb}{0,0.1,0.6}
\definecolor{dkgreen}{rgb}{0,0.35,0}
\definecolor{dkviolet}{rgb}{0.3,0,0.5}
\definecolor{dkred}{rgb}{0.5,0,0}
\definecolor{comment}{HTML}{444444}
\definecolor{keywd}{HTML}{8D00ED}
\definecolor{types}{HTML}{1F7B2F}
\definecolor{str}{HTML}{4070a0}
\definecolor{code-background}{gray}{0.8}
\definecolor{pragma}{HTML}{372A78}
\definecolor{num}{HTML}{40a070}
\definecolor{symb}{HTML}{000000}
\newcommand\symbmath[1]{{\ensuremath{\color{symb}#1}}}
\newcommand\typemath[1]{{\ensuremath{\uley{#1}}}}

\newcommand{\ord}[1]{{$\uley{\mathbb{N}}_{<\text{\texttt{#1}}}$}}
\lstdefinelanguage{Haskell}{
 morecomment=[l][\slshape\color{comment}]{--},
 morecomment=[n][\slshape\color{comment}]{\{-}{-\}},
 morecomment=[n][\bfseries\itshape]{\{-\#}{\#-\}},
 morestring=[b]\",
 stringstyle={\slshape},
 keywords={case,class,data,deriving,do,else,if,import,in,infixl,infixr,instance,let,infix,
	   module,of,primitive,then,type,where,family,newtype
 },
 numbers=left,
 stepnumber=1,
 keywordstyle={\bfseries},
 morekeywords=[2]{*,+,/,-,.,\%,!,?,=,<,>,$},
 keywordstyle=[2]{\ttfamily\bfseries},
 emph={[1] Bool,Char,Double,Either,Float,IO,Int,Maybe,Ordering,Rational,Ratio,ReadS,ShowS,String,%
	   Word8,InPacket,Sing,SNat,Show,Eq,Ord,Num,SingRep,Void,Monad,Field,Coeff,MyMat,%
           IsOrdPoly, CoeffRing, Ring, Abelian, Commutative, Group, Ideal, Unipol,Strategy,%
           OrdPoly,IsMonomialOrder,MOrder,Coeff,Arity,Lex,Grevlex,Revlex,Homogenised,%
           Additive,Monoidal,OrderedMonomial,Module,LeftModule,Scalar,Type,Polynomial,%
           MMatrix,Matrix,Mutable,Vector,ST,Entry,HPS,%
           Unital,Multiplicative,HomogOrder%
 },
 emphstyle={[1]\uley},
 emph={[2] EQ,False,GT,Just,LT,Left,Nothing,Right,True,Z,S,SS,SZ,sS,sZ
 },
 emphstyle={[2]\itshape},
 literate=
   {Nat}{{\typemath{\mathbb{N}}}}1 {Integer}{{\typemath{\mathbb{Z}}}}2
   {fromInteger}{{fromInteger}}{12}
   {KnownNat}{{KnownNat}}{10}
   {Rational}{{\typemath{\mathbb{Q}}}}2
   {::}{{\textcolor{symb}{:\!:}}}2
   {==}{{\symbmath{=}}}2
   {/=}{{\symbmath{\neq}}}2
   {forall}{{\ensuremath{\boldface{\forall}}}}1
   {*}{{\symbmath{\times}}}1
   {!*}{{\symbmath{\ltimes}}}2
   {+}{{\symbmath{+}}}1
   {++}{{\symbmath{{}^{\frown}}}}2
   {\%-}{{\symbmath{\ominus}}}2
   {P.++}{{\symbmath{{}^{\frown}}}}2
   {>}{{\symbmath{>}}}2
   {<}{{\symbmath{<}}}2
   {\\}{{\symbmath{\lambda}}}1 {~}{{\symbmath{\sim}}}1
   {:~:}{{\symbmath{\simeq}}}2
   {Fp}{{\typemath{\mathbb{F}_p}}}2
   {\\\\}{{\color{symb}\char`\\\char`\\}}2 {phi}{{\ensuremath{\varphi}}}1
   {->}{{\symbmath{\rightarrow}}}2 {>=}{{\symbmath{\geq}}}2 {<-}{{\symbmath{\leftarrow}}}2
   {`elem`}{{\symbmath{\in}}}2 {kappa}{{\ensuremath{\kappa}}}1 {===}{{\symbmath{\equiv}}}2
   {`IsSubsetOf`}{{\symbmath{\subseteq}}}2
   {==>}{{\symbmath{\Longrightarrow}}}{3}
   {<=}{{\symbmath{\leq}}}1 {=>}{{\symbmath{\Rightarrow}}}2  {=~=}{{\symbmath{\cong}}}2
   {\ .}{{\symbmath{\circ}}}3 {\ .\ }{{\symbmath{\circ}}}3 {:=:}{{\symbmath{\approx}}}2
   {\ ..}{{{\color{symb}\ .\!\!.}}}2
   {\ ...}{{\ensuremath{\ .\hspace{-.1ex}.\hspace{-.1ex}.}}}2
   {\ .*}{{\ \symbmath{\bullet}}}{2}
   {\ .||.}{{\ \symbmath{\vee}}}3 {\ .&&.}{{\symbmath{\ \wedge}}}2
   {\ .=}{{\ \symbmath{{:}{=}}}}4
   {\ .\%=}{{\ \symbmath{{:}{\Leftarrow}}}}4
   {<\$>}{{\symbmath{{<}\!{\text{\textdollar}}\!{>}}}}3
   {<*>}{{\symbmath{{<}\!\!\!\ast\!\!\!{>}}}}3
   {>>}{{\symbmath{\ll}}}1 {>>=}{{\symbmath{\gg\mkern-6.5mu=}}}2
   {=<<}{{\symbmath{=\mkern-6.5mu\ll}}}3
   {|}{{\symbmath{\mid}}}1
   {F\ }{{\typemath{\mathbb{F}\ }}}2
   {GF\ }{{\typemath{\mathbb{GF}\ }}}3
}
\lstdefinestyle{plain}{
 basicstyle=\ttfamily,
 identifierstyle=,
 commentstyle={\color{comment}},
 keywordstyle={\bfseries\color{keywd}},
 stringstyle=,
 keywordstyle=,
 keywords={},
 emphstyle={},
 texcl=false,
 mathescape=false,
 literate={},
 showspaces=false,
 showstringspaces=false,
 showtabs=false,
 breaklines=true,
 sensitive=true,
 captionpos=t,
 language={},
 frame=trbl,
 rulecolor={\color{white}},
 columns=[c]fixed,
 keepspaces=true,
}
\lstnewenvironment{code}[1][]{
  \lstset{basicstyle=\ttfamily, breaklines=false, #1}
}{}

\title{A Purely Functional Computer Algebra System Embedded in Haskell}
\author{Hiromi Ishii}
\institute{%
  Doctoral Program in Mathematics,\\%
  University of Tsukuba, Tsukuba, Ibaraki 305-8571, Japan\\%
  \email{h-ishii@math.tsukuba.ac.jp}%
}

\begin{document}
\maketitle
%#!luajitlatex -src-specials ms.tex

\begin{abstract}
 We demonstrate how methods in \emph{Functional Programming} can be used to implement a computer algebra system.
 As a proof-of-concept, we present the \texttt{computational-algebra} package.
 It is a computer algebra system implemented as an embedded domain-specific language in \emph{Haskell}, a purely functional programming language.
 Utilising methods in functional programming and prominent features of Haskell, this library achieves safety, composability, and correctness at the same time.
 To demonstrate the advantages of our approach, we have implemented advanced Gr\"{o}bner basis algorithms, such as Faug\`{e}re's $F_4$ and $F_5$, in a composable way.

 \vskip 1em
 \noindent{}\textbf{Keywords:} \mykeywords.
\end{abstract}

% Local Variables:
% mode: yatex
% TeX-master: "config.tex"
% End:

%#!luajitlatex -src-specials ms.tex

\section{Introduction}
In the last few decades, the area of computational algebra has grown larger.
Many algorithms have been proposed, and there have emerged plenty of computer algebra systems.
Such systems must achieve \emph{correctness}, \emph{composability} and \emph{safety} so that one can implement and examine new algorithms within them.
More specifically, we want to achieve the following goals:

\begin{description}
 \item[Composability] means that users can easily implement algorithms or mathematical objects so that they work seamlessly with existing features.
 \item[Safety] prevents users and implementors from  writing ``wrong'' code.
            For example, elements in different rings, e.g. $\Q[x,y,z]$ and $\Q[w,x,y]$, should be treated differently and must not directly be added.
            Also, it is convenient to have handy ways to convert, inject, or coerce such values.
 \item[Correctness] of algorithms, with respect to prescribed formal specifications, should be guaranteed with a high assurance.
\end{description}

We apply methods in the area of \emph{functional programming} to achieve these goals.
As a proof-of-concept, we present the \texttt{computational-algebra} package~\cite{computational-algebra}.
It is implemented as an embedded domain-specific language in the \emph{Haskell} Language \cite{Haskell}.
More precisely, we adopt the \emph{Glasgow Haskell Compiler} (GHC) \cite{GHCTeam:2018bq} as our hosting language.
We use GHC because: its \emph{type-system} allows us to build a safe and composable interface for computer algebra; \emph{lazy evaluation} enables us to treat infinite objects intuitively; \emph{declarative style} sometimes reduces a burden of writing mathematical programs; \emph{purity} permits a wide range of equational optimisation; and there is a plenty of libraries for functional methods, especially \emph{property-based testing}.
These methods are not widely adopted in this area; an exception is \emph{DoCon}~\cite{Mechveliani:2001fr}, a pioneering work combining Haskell and computer algebra.
Our system is designed with more emphasis on safety and correctness than DoCon, adding more ingredients.
Although we use a functional language, some methods in this paper are applicable in imperative languages.

This paper is organised as follows.
In \Cref{sec:type-system}, we discuss how the progressive type-system of GHC enables us to build a safe and expressive type-system for a computer algebra.
Then, in \Cref{sec:lightw-corr-prop}, we see how the method of \emph{property-based testing} can be applied to verify the correctness of algebraic programs in a lightweight and top-down manner.
To demonstrate the practical advantages of Haskell, \Cref{sec:examples} gives a brief description of the current implementations of the Hilbert-driven, $F_4$ and $F_5$ algorithms. We also take a simple benchmark there.
We summarise the paper and discuss related and future works in \Cref{sec:concl}.

In what follows, we use symbols in \Cref{tab:notation} in code fragments for readability.

\begin{table}[tbp]
 \caption{Symbols in Code Fragments\label{tab:notation}}
 \centering
  \newcolumntype{C}{>{\centering}p{0.1125\textwidth} }
  \newcolumntype{e}{p{0pt}}
  \setlength{\tabcolsep}{1pt}
  \begin{tabular}{@{\,}CC@{\hspace{6pt}}CC@{\hspace{6pt}}CC@{\hspace{6pt}}CCe@{\,}}
   \toprule
   Symbol & Code & Symbol & Code & Symbol & Code & Symbol & Code &\\
   \midrule
   {\typemath{\mathbb{N}}} & \lstinline[literate={}]!Nat! &
   {\typemath{\mathbb{Z}}} & \lstinline[literate={}]!Integer! &
   {\typemath{\mathbb{Q}}} & \lstinline[literate={}]!Rational! &
   {\typemath{\mathbb{F}_p}} & \lstinline[literate={}]!F p! &
   \\
   {\textcolor{symb}{::}} & \lstinline[literate={}]!::! &
   {\symbmath{=}} & \lstinline[literate={}]!==! &
   {\symbmath{\neq}} & \lstinline[literate={}]!/=! &
   {\symbmath{\lambda}\texttt{ }$\vec{\mathtt{x}}${ }\symbmath{\rightarrow}\texttt{ e}} & \lstinline[literate={}]!\!\ensuremath{\vec{\mathtt{x}}}\lstinline[literate=]! -> e! &
   \\
   {\symbmath{\times}} & \lstinline[literate={}]!*! &
   {\symbmath{\ltimes}} & \lstinline[literate={}]+!*+ &
   {\symbmath{{}^{\frown}}} & \lstinline[literate={}]!++! &
   {\symbmath{\ominus}} & \lstinline[literate={}]!%-! &
   \\
   {\symbmath{\simeq}} & \lstinline[literate={}]!:~:! &
   { \symbmath{\sim}} & \lstinline[literate={}]!~! &
   {\symbmath{\rightarrow}} & \lstinline[literate={}]!->! &
   { \symbmath{\leftarrow}} & \lstinline[literate={}]!<-! &
   \\
   \symbmath{\Longrightarrow} & \lstinline[literate={}]!==>!&
   \symbmath{\Rightarrow} & \lstinline[literate={}]!=>!&
   \symbmath{\symbmath{{:}{=}}} & \lstinline[literate={}]!.=!&
   \symbmath{{:}{\Leftarrow}} & \lstinline[literate={}]!.%=!&
   \\
   \symbmath{\subseteq} & {\verb+`Subset`+}&
   \symbmath{\leq} & \lstinline[literate={}]!<=!&
   \symbmath{\circ} & \lstinline[literate={}]!.!&
   \symbmath{\wedge} & \lstinline[literate={}]!.&&.!&
   \\
   \symbmath{\bullet} & \lstinline[literate={}]!.*!&
   \symbmath{{<}\!{\text{\textdollar}}\!{>}} & \lstinline[literate={}]!<$>!&
   \symbmath{{<}\!\!\!\ast\!\!\!{>}} & \lstinline[literate={}]!<*>!&
   {\ensuremath{\boldface{\forall}}} & \lstinline[literate={}]!forall! &
   \\
   \bottomrule
  \end{tabular}
\end{table}

% \subsection{Brief Explanation of Haskell}
% \emph{Haskell} is a statically typed purely functional programming language which has been evolving for the decades.
% The \emph{purity} means that the every expression of Haskell does not include any side-effects, that is, returns the same result when given the same inputs.
% Haskell employs the concept of \emph{monads} to encapsulate the side-effects while maintaining the purity.
% As a mental model, every effectful computation can be regarded  first  to be constructed as an abstract instruction, and then the compiler finally interprets it to the real I/O.
% It is this purity that enables Haskell to enjoy the lightweight parallelism and rewriting rules.

% Local Variables:
% mode: yatex
% TeX-master: "config.tex"
% End:

%  LocalWords:  lightw tex Fp GF

%#!luajitlatex -src-specials ms.tex

\section{Type System for Safety and Composability}\label{sec:type-system}
In this section, we will see how the progressive type-level functionalities of GHC can be exploited to construct a safe, composable and flexible type-system for a computer algebra system.
There are several existing works on type-systems for computer algebra, such as in Java and Scala~\cite{Kredel:2010jy,Jolly:2013rt}, and DoCon.
However, none of them achieves the same level of safety and composability as our approach, which utilises the power of \emph{dependent types} and \emph{type-level functions}.

\subsection{Type Classes to Encode Algebraic Hierarchy}
We use \emph{type-classes}, an ad-hoc polymorphism mechanism in Haskell, to encode an algebraic hierarchy.
This idea is not particularly new (for example, see Mechveliani~\cite{Mechveliani:2001fr} or Jolly~\cite{Jolly:2013rt}), and we build our system on top of the existing \texttt{algebra} package~\cite{Kmett:2011kb}, which provides a fine-grained abstract algebraic hierarchy.

\begin{flcode}
\begin{code}
class Additive a where(*\label{line:add-head}*)
  (+) :: a -> a -> a(*\label{line:plus}*)
class Additive a => Monoidal a where(*\label{line:mon-head}*)
  zero :: a(*\label{line:zero}*)
class Monoidal a => Group a where
  negate :: a -> a
\end{code}

\caption{Group structure, coded in the \texttt{algebra} package}
 \label{fig:ring-algebra}
\end{flcode}

\Cref{fig:ring-algebra} illustrates a simplified version of the algebraic hierarchy up to \lstinline{Group} provided by the \texttt{algebra} package.
Each statement between \lstinline{class} or \lstinline{=>} and \lstinline{where}, such as \lstinline{Additive a} or \lstinline{Monoidal a}, expresses the constraint for types.
For example, \Cref{line:add-head,line:plus} express ``a type \lstinline{a} is \lstinline{Additive} if it is endowed with a binary operation \lstinline{+}'',
and \Cref{line:mon-head,line:zero} that ``a type \lstinline{a} is \lstinline{Monoidal} if it is \lstinline{Additive} and has a distinguished element called \lstinline{zero}''.

Note that, none of these requires the ``proof'' of algebraic axioms.
Hence, one can accidentally write a non-associative \lstinline{Additive}-instance, or non-distributive \lstinline{Ring}-instance\footnote{Indeed, one can use \emph{dependent types}, described in the next subsection, to require such proofs. However, this is too heavy for the small outcome, and does not currently work for primitive types.}.
This sounds rather ``unsafe'', and we will see how this could be addressed reasonably in \Cref{sec:lightw-corr-prop}.

\subsection{Classes for Polynomials and Dependent Types}
\begin{flcode}[tbp]
 \begin{code}
 class (Module (Coeff poly) poly, Commutative poly, Ring poly,
        CoeffRing (Coeff poly), IsMonomialOrder (MOrder poly))
    => IsOrdPoly poly where
   type Arity  poly :: Nat
   type MOrder poly :: Type
   type Coeff  poly :: Type
   liftMap :: (Module (Scalar (Coeff poly)) alg, Ring alg)
           => ((*\ord{Arity poly}*) -> alg) -> poly -> alg
   leadTerm :: poly -> (Coeff poly, OrdMonom (MOrder poly) n)
   ...\end{code}
 \caption{A type-class for polynomials}
 \label{code:polyns}
\end{flcode}
Expressing algebraic hierarchy using type-class hierarchy, or class inheritance, is not so new and they are already implemented in DoCon or JAS.
However, these systems lack a functionality to distinguish the arity of polynomials or the denominator of a quotient ring.
In particular, DoCon uses sample arguments to indicate such parameters, and they cannot be checked at compile-time.
To overcome these restrictions, we use \emph{Dependent Types}.

For example, \Cref{code:polyns} presents the simplified definition of the class \lstinline{IsOrdPoly} for polynomials.
We provide an abstract class for polynomials, not just an implementation, to enable users to choose appropriate internal representations fitting their use-cases.

The class definition includes not only functions, but also \emph{associated types}, or \emph{type-level functions}: \lstinline{Arity}, \lstinline{MOrder} and \lstinline{Coeff}.
Respectively, they correspond to the number of variables, the monomial ordering and the coefficient ring.

Note that \lstinline{liftMap} corresponds to the universality of the polynomial ring $R[X_1, \dots, X_n]$; i.e.\ the free associative commutative $R$-algebra over $\set{1, \dots, n}$.
In theory, this function suffices to characterise the polynomial ring.
However, for the sake of efficiency, we also include some other operations in the definition.

\begin{flcode}[tbp]
 \begin{code}
 instance (IsMonomialOrder ord, CoeffRing r)
       => IsOrdPoly (OrdPoly r ord n) where
   type Arity  (OrdPoly r ord n) = n
   type MOrder (OrdPoly r ord n) = ord
   type Coeff  (OrdPoly r ord n) = r
   ...

 f :: OrdPoly Rational Grevlex 3(*\label{line:f-sig}*)
 f = let [x,y,z] = vars in x ^ 2 * y + 3 * x + z + 1

 instance (CoeffRing r) => IsOrdPoly (Unipol r) where
   type Arity  (OrdPoly r ord n) = 1(*\label{line:ordpol-arity}*)
   type MOrder (OrdPoly r ord n) = Lex
   type Coeff  (OrdPoly r ord n) = r
   ...
\end{code}
 \caption{Examples for polynomial instances}
 \label{code:polyn-insts}
\end{flcode}

\Cref{code:polyn-insts} shows example instance definitions for the standard multivariate and univariate polynomial ring types.
Note that, in \Cref{line:f-sig,line:ordpol-arity}, number literal \emph{expressions} $1$ and $3$ occur in \emph{type} contexts.
Types depending on expressions are called \emph{Dependent Types} in type theory.
GHC supports them via the \emph{Promoted Data-types} language extension~\cite{Yorgey:2012} since version 7.4.
Our library heavily uses this functionality, and achieves the type-safety preventing users from unintendedly confusing elements from different rings.

\subsection{Proofs in Dependent Types and Type-driven Casting Function}\label{sec:proof-depend-types}
\begin{flcode}
 \begin{code}
 convPoly :: (Coeff r ~ Coeff r', MOrder r ~ MOrder r', 
              Arity r ~ Arity r')
          => r -> r'
 injVars :: (Arity r <= Arity r', Coeff r ~ Coeff r')
         => r -> r'
 injVarsOffset :: (n + Arity r <= Arity r', Coeff r ~ Coeff r')
               => Sing n -> r -> r'
\end{code}
 \caption{Various casting function, with simplified type-signatures}
 \label{code:casts}
\end{flcode}
In theory, we can use \lstinline{liftMap} to cast between any elements of  ``compatible'' polynomial rings.
To reduce the burden to write boilerplate casting functions, our library comes with smart functions, as shown in \Cref{code:casts}.
The \texttt{convPoly} function maps a polynomial into one with the same setting but different representation; e.g. \lstinline{OrdPoly Rational Lex 1} into \lstinline{Unipol Rational}.
The next \lstinline{injVars} function maps an element of $R[X_1, \dots, X_n]$ into another polynomial ring with the same coefficient ring, but with more number of variables, e.g.\ $R[X_1, \dots, X_{n+m}]$, regardless of ordering.
For example, it maps \lstinline{Unipol Rational} into \lstinline{OrdPoly Rational Grevelx 3}.
Then, \lstinline{injVarsOffset} is a variant of \lstinline{injVars} which maps variables with offset; for example,
\begin{code}
 injVarsOffset [sn|3|] :: Unipol Rational -> Polynomial Rational 5
\end{code}
maps $\Q[X]$ into $\Q[X_0, \dots, X_4]$ with $X \mapsto X_3$.
Here, \lstinline{[sn|3|]} is called a \emph{singleton} for the type-level natural number \lstinline{3}, first introduced by Eisenberg et al.~\cite{Eisenberg:2012}.
More precisely, for any \emph{type-level natural} \lstinline{n}, there is the unique \emph{expression} \lstinline{sing :: Sing n} and we can use it as a tag for type-level arguments.

To work with type-level naturals, we sometimes have to \emph{prove} some constraints.
For example, suppose we want to write a variant of \lstinline{injVars} mapping variables to \emph{the end of} those of the target polynomial ring, instead of \emph{the beginning}.
We might first write it as follows:
\begin{code}
injVarsAtEnd :: (Arity r <= Arity r', Coeff r ~ Coeff r')
             => r -> r'
injVarsAtEnd =
  let sn = sing :: Sing (Arity r)
      sm = sing :: Sing (Arity r')
  in injVarsOffset (sm %- sn) -- Errors!
\end{code}
However, GHC cannot see \lstinline{Arity r' - Arity r + Arity r <= Arity r'}.
Although this constraint is rather clear to us, we have to give the compiler its proof.
We have developed the \texttt{type-natural} package \cite{type-natural} which includes typical ``lemmas''.
For example, we can use the \lstinline{minusPlus} lemma to fix this:
\begin{code}
-- From type-natural:
minusPlus :: Sing n -> Sing m
          -> IsTrue (m <= n) -> ((n - m) + m) :~: n

injVarsAtEnd :: (Arity r <= Arity r', Coeff r ~ Coeff r')
             => r -> r'
injVarsAtEnd =
  let sn = sing :: Sing (Arity r)
      sm = sing :: Sing (Arity r')
  in withRefl (minusPlus sm sn Witness) $
     injVarsOffset (sm %- sn)
\end{code}
Since giving such a proof each time is rather tedious, we can use type-checker plugins to let the compiler try to prove constraints automatically.
In particular, the author developed the \texttt{ghc-typelits-presburger} plugin~\cite{ISHII:2017nq} to resolve propositions in Presburger arithmetic at compile time.

Our library also provides the \lstinline{LabPoly} type, which converts existing polynomial types into ``\emph{labelled}'' ones.
For example, one can write as follows:
\begin{code}
f :: LabPoly (Polynomial Rational 3) '["x", "y", "z"]
f = 5 * #x ^ 2 * #y ^ 3 - #y * #z + 1
\end{code}
This relies on the \texttt{DataKinds} and \texttt{OverloadedLabels} language extensions of GHC.
GHC's type system is strong enough to reject illegal terms and types, such as \lstinline{#w :: LabPoly (Unipol Rational) '["a"]} ($w$ is not listed as a variable) or \lstinline{LabPoly (Polynomial Rational 3) '["x", "y", "x"]} (the variable $x$ occurs twice).
Using the type-level information, one can invoke the canonical inclusion maps naturally as follows:
\begin{code}
f :: LabPoly' Rational Grevlex '["x", "y", "z"]
f = #x * #y * #z + 2 * #y - 3  * #z * #x + 1
g :: LabPoly' Rational Lex '["w", "z", "y", "u", "x"]
g = canonicalMap f

-- Where:
canonicalMap :: (xs `IsSubsetOf` ys, Wraps xs poly, Wraps ys poly',
                 IsPolynomial poly, IsPolynomial poly',
                 Coeff poly ~ Coeff poly')
             => LabPoly poly xs -> LabPoly poly' ys
\end{code}

\subsection{Optimising Casting Functions with Rewriting Rules}
Since the casting functions are implemented generically, they sometimes introduce unnecessary overhead.
For example, if one uses \lstinline{injVars} with the \emph{same} source and target types, it should just be the identity function.
Fortunately, we can use the type-safe \emph{Rewriting Rule} functionality of GHC to achieve this:
\begin{code}
{-# RULES "injVars/identity" injVars = id #-}
\end{code}
Each rewriting rule fires at compile-time, if there is a term matching the left-hand side of the rule and having the same type as the right-hand side.

In Haskell, it suffices just to consider algebraic laws to write down custom rewriting rules.
This is due to the \emph{purity} of Haskell.
That is, every expression in Haskell is pure, in a sense that they evaluate to the same result when given the same arguments.
Note that this does not mean that Haskell cannot treat values with side-effects; indeed, the type-system of Haskell distinguishes pure and impure values at type-level, and one can treat impure operations without violating purity as a whole.
The trick behind this situation is to describe side-effects as some kind of abstract instructions, instead of treating impure values directly.
Hence, for example, duplicating the same term does not make any difference in its meaning, provided that it is algebraically correct.
Such a rewriting rule is used extensively in Haskell.
For example, Stream Fusion \cite{Coutts:2007uq} uses them to eliminate unnecessary intermediate expressions and fuse complicated functions into efficient one-path constructions.
Yet, DoCon did not do any optimisation using rewriting rules.

In our library, we also use rewriting rules to remove idempotent applications such as ``grading'' a monomial ordering twice, e.g:
\begin{code}
{-# RULES "graded/graded" forall ord. 
  graded (graded ord) = graded ord #-}
\end{code}

\subsection{Notes on applicability in imperative languages}
The safety we achieved in this section cannot be achieved at compile-time without dependent types and type-level functions.
Existing works using type-classes or class inheritance to encode algebraic hierarchy, such as JAS or DoCon, lack this level of safety.
In theory, one can achieve the same level of safety even in a statically-typed \emph{imperative} language, if it supports a kind of dependent types.
For example, in C\texttt{++}, templates with non-type arguments can be used to simulate dependent types.
On the other hand, in Java, Generics do not allow non-type arguments and we need to mimic Peano numerals with classes.
In either case, it requires much effort to prove the properties of naturals within them, because they lack dedicated support for type-level naturals or type-checker plugins.

On the other hand, to make use of rewriting rules, we need purity as discussed above.

% Local Variables:
% mode: yatex
% TeX-master: "ms.tex"
% End:

%  LocalWords:  fromInteger tbp Coeff CoeffRing IsMonomialOrder MOrder
%  LocalWords:  IsOrdPoly Arity liftMap alg leadTerm OrderedMonomial sn
%  LocalWords:  htbp ord OrdPoly Grevlex convPoly
%  LocalWords:  injVars injVarsOffset

%#!luajitlatex -src-specials ms.tex

\section{Lightweight Correctness: Property-based Testing}\label{sec:lightw-corr-prop}

\subsection{Property-based testing introduced}
In this section, we will address the correctness issue, in a top-down, or \emph{lightweight}  manner.
Especially, we apply the method of \emph{property-based testing} \cite{Claessen:2000cz} to verify the correctness of our implementation.
The idea is that one specifies the formal properties that the implemented algorithms and types must satisfy, and checks if they hold by testing them against randomly or exhaustively generated inputs.
Although it is not as rigorous as a theorem proving, it still gives a guarantee of the correctness at high assurance, after repeating tests time after time.

\begin{flcode}
\begin{code}
prop_division :: Rational -> Property(*\label{line:prop_div}*)
prop_division q =
    q /= 0 ==> (recip q * q == 1 .&&. q * recip q == 1)
  .&&. q * 1 == q .&&. 1 * q == q(*\label{line:prop_div-end}*)

prop_passesSTest n =
  forAll (idealOfArity n) $ \ ideal ->
  let gs = calcGroebnerBasis (toIdeal ideal)
  in all (isZero . (`modPoly` gs))
         [sPoly f g | f <- gs, g <- gs, f /= g]
\end{code}
 \caption{Formal Specification of Algebraic Programs}
 \label{code:specs}
\end{flcode}

\Cref{code:specs} presents the example specifications for algebraic programs.
In Lines \ref{line:prop_div} through \ref{line:prop_div-end}, \lstinline{prop_division} states that the implementation of \lstinline{Rational} must satisfy the axioms of division ring.
The \lstinline{prop_passesSTest} function demand the result of \lstinline{calcGroebnerBasis} to pass the $S$-test.
The tester accepts the specifications above, generates a specified number of inputs (default: $100$) and tests against them.
If all the inputs satisfy the specifications, it successfully halts; otherwise, it reports counterexamples, which is useful while debugging.

\subsection{Discussion}
There are several libraries for property-based testing adopting different strategies to generate inputs.
For example, QuickCheck~\cite{Claessen:2000cz} generates inputs randomly, while SmallCheck~\cite{Runciman:2008jb} exhaustively enumerates inputs in the depth-increasing order.
Even though there are other implementations of property-based testers in languages other than Haskell \cite{Hypothesis:2018rp}, it does not seem that it is applied in existing systems, such as Singular~\cite{Greuel:2007:SIC:1557288}, JAS or DoCon.

By its \emph{generative} nature, property-based testing has several drawbacks and pitfalls.
First, evidently, it cannot assure the validity as rigorously as the \emph{formal theorem proving}, unless the input space is finite.
There are several pieces of research that combine formal theorem proving and computational algebra to rigorously certify correctness of implementations (for example, \cite{Meshveliani:2018fv,Coquand:1999yo}).
These first formalise the theory of Gr\"{o}bner basis in the constructive type-theory.
Then, execute them within the host theorem proving language, or extract the program into other languages.
However, by its nature, this approach requires everything to be proven formally.
It is not so easy a task to prove the correctness of every part of a program, even with help from automatic provers.
Even if one manages to finish the proof of the validity of some algorithm, when one wants to optimise it afterwards, then one must prove the ``equivalence'' or validity of that optimisation.
Moreover, it is sometimes the case that the validity, or even termination, of the algorithm remains unknown when it is implemented; e.g.\ the correctness and termination of Fauger\`{e}'s $F_5$ \cite{Faugere:2002:NEA:780506.780516} are proven very recently~\cite{Pan:2013oa}.
Furthermore, there is an obvious restriction that we can extract programs only into the languages supported by the theorem prover.
We consider these conditions too restrictive, and decided to adopt theorem proving only in trivial arity arithmetic.

Secondly, if the algorithm has a bad time complexity, property-based tests can easily explode.
Specifically, since Gr\"{o}bner bases have double-exponential worst time complexity, randomly generated input can take much time to be processed.
One might reduce the burden by combining randomised and enumerative generation strategies carefully, but there is still a possibility that there are small inputs which take much time.
To avoid such a circumstance, one can reduce the number of inputs, however it also reduces the assurance of validity.

Finally, they are not so good at treating \emph{existential properties}.
Although Small\nolinebreak Check provides the existential quantifier in its vocabulary, it just tries to find solutions up to a prescribed depth.
If solutions are relatively ``larger'' than its inputs, this results in \emph{false-negative} failures.
For example, one can write the following specification that demands each element of the result of \lstinline{calcGroebnerBasis} to be a member of the original ideal, however it does not work as expected:
\begin{code}
prop_gbInc ideal =
  let j = calcGroebnerBasis ideal
  in exists $ \ cs ->
     and (zipWith (\ f gs -> f == dot ideal gs) j cs)
\end{code}
In the above, \lstinline{dot i g} denotes the ``dot-product''.
As a workaround, we currently combine inter-process communication with property-based testing.
More specifically, we invoke a reliable existing implementation, such as SINGULAR, inside the spec as follows:
\begin{code}
prop_gbInc = forAll arbitrary $ \ i -> monadicIO $ do
  let gs = calcGroebnerBasis i
  is <- evalSingularIdealWith [] [] $
        funE "reduce" [
          idealE gs, funE "groebner" [idealE i]]
  return $ all isZero is
\end{code}
Thus, if the existential property in question is decidable and has an existing reliable implementation,
then it might be better to call it inside specifications.

% Local Variables:
% mode: yatex
% TeX-master: "config.tex"
% End:

%#!luajitlatex -src-specials ms.tex

\section{Case Study: Implementing the Hilbert-driven, $F_4$ and $F_5$ algorithms for calculating Gr\"{o}bner bases}
\label{sec:examples}
In this section, we will focus on three algorithms as case-studies: the Hilbert-driven, $F_4$ and $F_5$ algorithms.
Firstly, we demonstrate the power of laziness and parallelism by the Hilbert-driven algorithm.
Then by the $F_4$ interface, we illustrate the practical example of composability.
Finally, we skim through the simplified version of the main routine of $F_5$ and see how imperative programming with mutable states can be written purely in Haskell.
For our purpose, we will discuss only a fragment of implementations that elucidates the advantages of Haskell, rather than the entire implementation and theoretical details.

\subsection{Homogenisation and Hilbert-driven basis conversion}
\begin{flcode}
\begin{code}
data Homogenised poly
instance IsOrdPoly poly => IsOrdPoly (Homogenised poly) where
  type Arity  (Homogenised poly) = 1 + Arity poly
  type MOrder (Homogenised poly) = HomogOrder (MOrder poly)
  type Coeff  (Homogenised poly) = Coeff poly
  ...
homogenise   :: IsOrdPoly poly => poly -> Homogenised poly
unhomogenise :: IsOrdPoly poly => Homogenised poly -> poly

calcGBViaHomog :: (Field (Coeff poly), IsOrdPoly poly)
               => (forall r. (Field (Coeff r), IsOrdPoly r)
                       => Ideal r -> [r])
               -> Ideal poly -> [poly]
calcGBViaHomog calc i
  | all isHomogeneous i = (*\ovalcode{fill=white,draw=black}{calc}*) i(*\label{line:homog-case}*)
  | otherwise = map unhomogenise ((*\ovalcode{fill=black!25}{calc}*) (fmap homogenise i))(*\label{line:nonhomog-case}*)
\end{code}
 \caption{Basic API for homogenisation}\label{code:homog-hilb-driv}
\end{flcode}

Homogenisation is a powerful tool in Gr\"{o}bner basis computation.
If $I \subseteq k[\mathbf{X}]$ is a non-homogeneous ideal and $\bar{I} \subseteq k[x, \mathbf{X}]$ its homogenisation, then one can get a Gr\"{o}bner basis for $I$ by unhomogenising the Gr\"{o}bner basis $\bar{G}$ for $\bar{I}$ w.r.t.\ a suitably induced monomial ordering.
In this way, any Gr\"{o}bner basis algorithm for homogeneous ideals can be converted into one for non-homogeneous ones.

\Cref{code:homog-hilb-driv} is an API for these operations.
The type \lstinline{Homogenised poly} represents polynomials obtained by homogenising polynomials of type \lstinline{poly}.
Then \lstinline{calcGBViaHomog calc i} first checks if the input \lstinline{i} is homogeneous.
If it is so, then it applies the argument \ovalcode{fill=white,draw=black}{calc} to its input directly (\Cref{line:homog-case}); otherwise, it first homogenises the input, applies \ovalcode{fill=black!25}{calc}, and then unhomogenises it to get the final result (\Cref{line:nonhomog-case}).
Note that, though it uses the same term \lstinline{calc} in both cases, they have different types.
In the first case, since it just feeds an input directly, \ovalcode{fill=white,draw=black}{calc} has type \lstinline{Ideal poly -> [poly]}.
On the other hand, in the non-homogeneous case, it is applied \emph{after} homogenisation, hence it is of type \lstinline{Ideal (Homogenised poly) -> [Homogenised poly]}.
Thus, \lstinline{calcGBViaHomog} takes a \emph{polymorphic function} as its first argument and this is why we have \lstinline{forall} inside the type of the first argument.
Such a nested polymorphic type is called a \emph{rank $n$ polymorphic type}, and it is supported by GHC's \texttt{RankNTypes} language extension\footnote{This can be achieved in object-oriented language with subtyping and Generics.}.

\begin{flcode}
\begin{code}
data HPS n = HPS { taylor :: [Integer], hpsNumerator :: Unipol Integer }

instance Eq (HPS a) where
  (==) = (==) `on` hpsNumerator
instance Additive (HPS n) where
  HPS cs f + HPS ds g = HPS (zipWith (+) cs ds) (f + g)
instance LeftModule (Unipol Integer) (HPS n) where
  f .* HPS cs g = HPS (conv (taylor f ++ repeat 0) cs) (f * g) 

conv :: [Integer] -> [Integer] -> [Integer]
conv (x : xs) (y : ys) =
  let parSum a b c = a (*\ovalcode[1pt]{fill=black!25}{`par`}*) b (*\ovalcode[1pt]{fill=black!25}{`par`}*) c (*\ovalcode[1pt]{fill=white,draw=black}{`seq`}*) (a + b + c) in(*\label{line:conv}*)
  x * y :
   zipWith3 parSum (map (x*) ys) (map (y*) xs) (0 : conv xs ys)
\end{code}
\caption{Data-type of and operations on Hilbert--Poincar\'{e} series}
\label{code:hps}
\end{flcode}
For example, one can use the so-called \emph{Hilbert-driven algorithm} as the first argument to \lstinline{calcGBViaHomog}.
It first computes a Gr\"{o}bner basis w.r.t.\ a lighter monomial ordering, compute the Hilbert--Poincar\'{e} series (HPS) with it and use it to compute Gr\"{o}bner basis w.r.t.\ the heavier ordering.
In this procedure, we need the following operations on HPS:
Equality test on HPS's, $n$\textsuperscript{th} Taylor coefficient of the given HPS, and the $\Z[X]$-module operation on HPS.
\Cref{code:hps} illustrates such an interface for HPS.
For equality test, we use the numerator \lstinline{hpsNumerator} of the closed form, and an \emph{infinite list}  \lstinline{taylor} maintains Taylor coefficients.
By the \emph{lazy} nature of Haskell, we can intuitively treat infinite lists and write a convolution on them.
In \Cref{line:conv}, \ovalcode{fill=black!25}{par} and \ovalcode{fill=white,draw=black}{seq} specify the \emph{evaluation strategy}.
In brief, expressions \lstinline{x} and \lstinline{y} in ``\lstinline{x }\ovalcode[1pt]{fill=black!25}{`par`}\lstinline{ y}'' (resp.  \ovalcode{fill=white,draw=black}{seq}) are evaluated \emph{parallelly} (resp.\ \emph{sequentially}).
Since every expression is pure in Haskell, we can safely take advantage of parallelism, without a possibility of changing results.

\subsection{A composable implementation of $F_4$}
\begin{flcode}[htbp]
\begin{code}
class MMatrix mat a where(*\label{line:mmat-head}*)
  fromRows :: [Vector a] -> ST s (mat s a)
  scaleRow :: Multiplicative a => Int -> a -> mat s a -> ST s ()
  ...

class MMatrix (Mutable mat) a => Matrix mat a where(*\label{line:mat-head}*)
  type Mutable mat :: (*$\star$*) -> (*$\star$*)
  freeze :: Mutable mat s a -> ST s (mat a)
  ...
  gaussReduction :: Field a => mat a -> mat a

type Strategy f w = f -> f -> w
f4 :: (Ord w, IsOrdPoly poly, Field (Coeff poly),
       Matrix mat (Coeff poly))
   => proxy mat -> Strategy poly w -> Ideal poly -> [poly](*\label{line:f4}*)
\end{code}
 \caption{Matrix classes and the $F_4$ function}
 \label{code:mat-f4}
\end{flcode}
$F_4$ is one of the most efficient algorithms for Gr\"{o}bner basis computation and introduced by Faug\`{e}re~\cite{Faugere:1999}.
Briefly, $F_4$ reduces more than two polynomials at once, replacing $S$-polynomial remaindering in the Buchberger Algorithm with the \emph{Gaussian elimination} of the matrices.
This means that the efficiency of $F_4$ reduces to that of Gaussian elimination and the internal representation of matrices.
Thus, it is useful if we can easily switch internal representations and elimination algorithms.
For this purpose, we provide type-classes for mutable and immutable matrices which admit row operations and a dedicated Gaussian elimination.
\Cref{code:mat-f4} demonstrates the interface for immutable and mutable matrices (\lstinline{Matrix} and \lstinline{MMatrix}) and the type signature of our $F_4$ implementation (\lstinline{f4}).
In \Cref{line:mmat-head,line:mat-head}, the last type argument \lstinline{a} of \lstinline{Matrix} and \lstinline{MMatrix} corresponds to the type of coefficients.
Note that, one can give different instance definitions for the same \lstinline{mat} but different coefficient types \lstinline{a}.
For example, one can implement efficient Gaussian elimination on $\mathbb{F}_p$ for \lstinline{Matrix Mat Fp}, and then use it in the definition of \lstinline{Matrix Mat Rational}, with the Hensel lifting or Chinese remaindering.

In \Cref{line:f4}, the first argument of \lstinline{f4} of type \lstinline{proxy mat} specifies the internal representation \lstinline{mat} of matrices.
In addition,  \lstinline{f4} takes a \emph{selection strategy} as the second argument.
Here, the selection strategy is abstracted as a weighting function to some ordered types, and we store intermediate polynomials in a heap and select all the polynomials with the minimum weight at each iteration.

\subsection{The $F_5$ algorithm}
\begin{flcode}[htbp]
\begin{code}
f5 :: (Field (Coeff pol), IsOrdPoly pol)
   => Vector pol -> [(Vector pol, pol)]
f5 (map monoize -> i0) = runST $ do
  let n = length i0
  gs <- newSTRef []
  ps <- newSTRef $ H.fromList [ basis n i | i <- [0..n-1]]
  syzs <- newSTRef
    [ sVec (i0 ! m) (i0 ! n) | m <- [0..n-1], n <- [0..j-1] ]
  whileJust_ (H.viewMin <$> readSTRef ps) $
  \ (Entry sig g, ps') -> do
    ps .= ps'
    (gs0, ss0) <- (,) <$> readSTRef gs <*> readSTRef syzs
    unless (standardCriterion sig ss0) $ do
      let (h, ph) = reduceSignature i0 g gs0
          h' = map ((*${\color{symb} \times}$*) injectCoeff (recip $ leadingCoeff ph)) h
      if isZero ph then syzs .%= (mkEntry h : )
        else do
        let adds = fromList $ mapMaybe (regSVec (ph, h')) gs0
        ps .%= H.union adds
        gs .%= ((monoize ph, mkEntry h') :)
  map (\ (p, Entry _ a) -> (a, p)) <$> readSTRef gs
\end{code}
 \caption{Main Routine of the $F_5$ Algorithm\label{code:F5}}
\end{flcode}

Finally, we present the simplified version of the main routine of Faug\`{e}re's $F_5$~\cite{Faugere:2002:NEA:780506.780516} (\Cref{code:F5}).
Readers may be surprised that the code looks much imperative.
This is made possible by the \emph{ST monad}~\cite{Launchbury:1994dk}, which encapsulates side-effects introduced by mutable states and prevents them from leaking outside.
We use a functional heap to choose the polynomial vectors with the least signature, demonstrating the fusion of functional and imperative styles.
\subsection{Benchmarks}
\begin{table}[h]
 \renewcommand{\arraystretch}{1.2}
 \centering

 \caption{Benchmark results (\si{ms})} \label{tab:bench}
  \newcolumntype{d}{%
   S[retain-zero-exponent = true,%
     table-format=1.3e+1%
   ]}
  \setlength{\tabcolsep}{5pt}
  \scriptsize
  \begin{tabular}[htbp]{lddddd}
   \toprule
   & {$I_1$ (Lex)}   & {$I_1$ (Grevlex)} & {$I_2$ (Lex)} & {$I_2$ (Grevlex)} & {$I_3$ (Grevlex)}\\
   \midrule          
   B      & 1.820e0  & 1.593e1  & 1.400e1  & 4.129e0  & 6.689e2
   \\
   DbyD   & 6.364e1  & 9.162e2  & 1.147e2  & 5.647e1  & 4.125e2
   \\
   Hilb   & 1.644e2  & 2.313e2  & 5.265e1  & 3.414e1  & 9.645e3
   \\
   $F_5$  & 1.851e0  & 4.314e2  & 7.129e0  & 2.648e0  & 1.290e3
   \\
   S(gr)  & 2.300e0  & 8.493e-1 & 2.651e0  & 8.210e-1 & 9.511e-1
   \\
   S(sba) & 2.279e-1 & 8.711e-1 & 2.343e-1 & 7.958e-1 & 1.541e-1
   \\
   \bottomrule
  \end{tabular}
 \begin{align*}
  I_1 \defeq \langle & 35  y^{4} - 30xy^{2} - 210y^{2}z + 3x^{2} + 30xz - 105z^{2} +140yt - 21u, \\
                     & 5xy^{3} - 140y^{3}z - 3x^{2}y + 45xyz - 420yz^{2} + 210y^{2}t -25xt + 70zt + 126yu\rangle\\
  I_2 \defeq \langle & w+x+y+z, wx+xy+yz+zw, wxy + xyz + yzw + zwx, wxyz - 1 \rangle\\
  I_3 \defeq \langle & x^{31} - x^{6} - x- y, x^{8} - z, x^{10} -t \rangle
 \end{align*}
 \vskip -2em
\end{table}
We also take a simple benchmark and the result is shown in \Cref{tab:bench} (examples are taken from Giovini et al.~\cite{Giovini:1991}).
This compares the algorithms implemented in our \texttt{computational-algebra} package and Singular.
The first four rows correspond to the alrorithms implemented in our library; i.e.\ the Buchberger algorithm optimised with syzygy and sugar strategy (B), the degree-by-degree algorithm for homogeneous ideals (DbyD), the Hilbert-driven algorithm (Hilb), and $F_5$. S(gr) and S(sba) stand for the \texttt{groebner} and \texttt{sba} functions in the Singular computer algebra system 4.0.3.
The complete source-code is available on GitHub~\cite{computational-algebra}\footnote{More specifically, we used the implementation in commit \href{https://github.com/konn/computational-algebra/tree/70e6e7b}{\texttt{70e6e7b}}.}.
The benchmark program is compiled with GHC 8.2.2 with flags \lstinline!-O2 -threaded -rtsopts -with-rtsopts=-N!,
and ran on an Intel Xeon E5-2690 at 2.90 GHz, RAM 128GB, Linux 3.16.0-4 (SMP), using 10 cores in parallel.
We used the Gauge framework to report the run-time of our library, and the \lstinline{rtimer} primitive for Singular.
For actual benchmark codes, see \url{http://bit.ly/hbench1} and \href{http://bit.ly/hbench2}{\texttt{hbench2}}.
Unfortunately, in our system, $F_4$ takes much more computing time, hence we did not include the result.
The results show that, among the algorithms implemented in our system, $F_5$ works fine in general, though it takes much time in some specific cases.
Nevertheless, there remains much room for improvement to compete with the state-of-the-art implementations such as Singular, although there is one case where our implementation is slightly faster than Singular's \texttt{groebner} function.

% Local Variables:
% mode: yatex
% TeX-master: "ms.tex"
% End:

%#!luajitlatex -src-specials ms.tex

\section{Conclusions}\label{sec:concl}
In this paper, we have demonstrated how we can adopt the methods developed in the area of functional programming to build a computer algebra system.
Some of these methods are also applicable in imperative languages.

In \Cref{sec:type-system}, we presented a type-system strong enough to detect algebraic errors at compile-time.
For example, our system can distinguish number of variables of polynomial rings at type-level thanks to dependent types.
It also enables us to automatically generate casting functions and we saw how their overhead can be reduced using rewriting rules.
As for type-systems for a computer algebra system, there are several existing works~\cite{Kredel:2010jy,Mechveliani:2001fr}.
However, these systems are not safe enough for discriminating variable arity at type-level and don't make use of rewriting rules.

In \Cref{sec:lightw-corr-prop}, we successfully applied the method of \emph{property-based testing} for verification of the implementation, which is lightweight compared to the existing theorem-prover based approach \cite{Coquand:1999yo,Meshveliani:2018fv}.
Although property-based testing is not as rigorous as theorem proving, it is lightweight and can be applied to algorithms not yet proven to be valid or terminate and available also for imperative languages.

We have seen that, in \Cref{sec:examples}, other features of Haskell, such as higher-order polymorphism, parallelism and laziness, can also be easily applied to computer algebra by actual examples.
Even though they are shown as fragments of code, we expect them to be convincing.

Since some of the methods in this paper, such as dependent types or property-based testing, are not limited to the functional paradigm, it might be interesting to investigate their applicability in the imperative settings.

From the viewpoint of efficiency, there are much to be done.
For example, efficiency of our current $F_4$ implementation is far inferior to that of the na\"{\i}ve Buchberger algorithm, and other algorithms are far much slower than state-of-the-art implementations such as Singular.
To optimise implementations, we can make more use of Rewriting Rules and efficient data structures.
Also, the parallelism must undoubtedly play an important role.
Fortunately, there are plenty of the parallel computation functionalities in Haskell, such as Regular Parallel Arrays \cite{Keller:2010zr} and \texttt{parallel} package \cite{Marlow:2010tw}, and another book by Marlow~\cite{Marlow:2013rc} on general topics in parallelism in Haskell.
Also, there is an existing work by Lobachev et al.~\cite{Lobachev:2010qe} on parallel symbolic computation in Eden, a dialect of Haskell with parallelism support.
Although Eden is retired, the methods introduced there might be helpful.

\subsection*{Acknowledgements}
The author would like to thank my supervisor, Prof. Akira Terui, for discussions, and to anonymous reviewers for helpful comments.
This research is supported by Grant-in-Aid for JSPS Research Fellow Number 17J00479, and partially by Grants-in-Aid for Scientific Research 16K05035.
This is a pre-print of an article published in ``Computer Algebra in Scientific Computing'' (2018). The final authenticated version is available online at: \url{https://doi.org/10.1007/978-3-319-99639-4}

% Local Variables:
% mode: yatex
% TeX-master: "config.tex"
% End:

\begin{sloppypar}
 \printbibliography
\end{sloppypar}
\end{document}